\documentclass[a4paper]{article}

\usepackage[pages=all, color=black, position={current page.south}, placement=bottom, scale=1, opacity=1, vshift=5mm]{background}
\SetBgContents{
	\tt Bhowmik et al. 2023: Manuscript submitted to Journal
}      

\usepackage[margin=1in]{geometry} 

\usepackage{amsmath}
\usepackage{amsthm}
\usepackage{amssymb}

\usepackage[utf8]{inputenc}
\usepackage{hyperref}
\hypersetup{
	unicode,
	pdfauthor={Author One, Author Two, Author Three},
	pdftitle={A simple article template},
	pdfsubject={A simple article template},
	pdfkeywords={article, template, simple},
	pdfproducer={LaTeX},
	pdfcreator={pdflatex}
}

\usepackage[sort&compress,numbers,square]{natbib}
\theoremstyle{plain}

\theoremstyle{definition}

\usepackage{graphicx, color}
\graphicspath{{fig/}}

\usepackage{algorithm, algpseudocode} 
\usepackage{mathrsfs} 

\usepackage{lipsum}

\title{Role of modified cloud microphysics parameterization in coupled climate model for studying ISM rainfall: small-scale cloud model and climate model work better together}
\author{Moumita Bhowmik$^1$\thanks{Corresponding Author: moumita.bhowmik@tropmet.res.in} \and Anupam Hazra$^1$\thanks{Corresponding Author: hazra@tropmet.res.in} \and Ankur Srivastava$^1$ \and Dipjyoti Mudiar$^1$ \and Hemantkumar S. Chaudhari$^1$ \and  Suryachandra A. Rao$^1$ \and Lian-Ping Wang$^2$}

\date{
	$^1$Indian Institute of Tropical Meteorology, Ministry of Earth Sciences, Pune, India \\ %
	$^2$Center for Complex Flows and Soft Matter Research, Department of Mechanics and Aerospace Engineering, Southern University of Science and Technology, China \\ [2ex]%
}

\begin{document}
	\maketitle
	
	\begin{abstract}
		An unresolved problem of present generation coupled climate models is the realistic distribution of rainfall over Indian monsoon region, which is also related to the persistent dry bias over Indian land mass. Therefore, quantitative prediction of the intensity of rainfall events has remained a challenge for the state-of-the-art global coupled models. Guided by the observation, it is hypothesized that insufficient growth of cloud droplets and processes responsible for the cloud to rain water conversion are key components to distinguish between shallow to convective clouds. The new diffusional growth rates and relative dispersion based ‘autoconversion’ from the Eulerian-Lagrangian particle-by-particle based small-scale model provide a pathway to revisit the parameterizations in climate models for monsoon clouds. The realistic information of cloud drop size distribution is incorporated in the microphysical parameterization scheme of climate model. Two sensitivity simulations are conducted using coupled forecast system (CFSv2) model.  
        When our physically based small-scale derived modified parameterization is used, a coupled climate model simulates the probability distribution (PDF) of rainfall and accompanying specific humidity, liquid water content, and outgoing long-wave radiation (OLR) with increasing accuracy. The improved simulation of rainfall PDF appears to have been aided by much improved simulation of OLR and resulted better simulation of the ISM rainfall.

	\end{abstract}

	
	\section{Introduction}
\begin{figure}[htbp]
\centerline{\includegraphics[height=6cm,width=12cm]{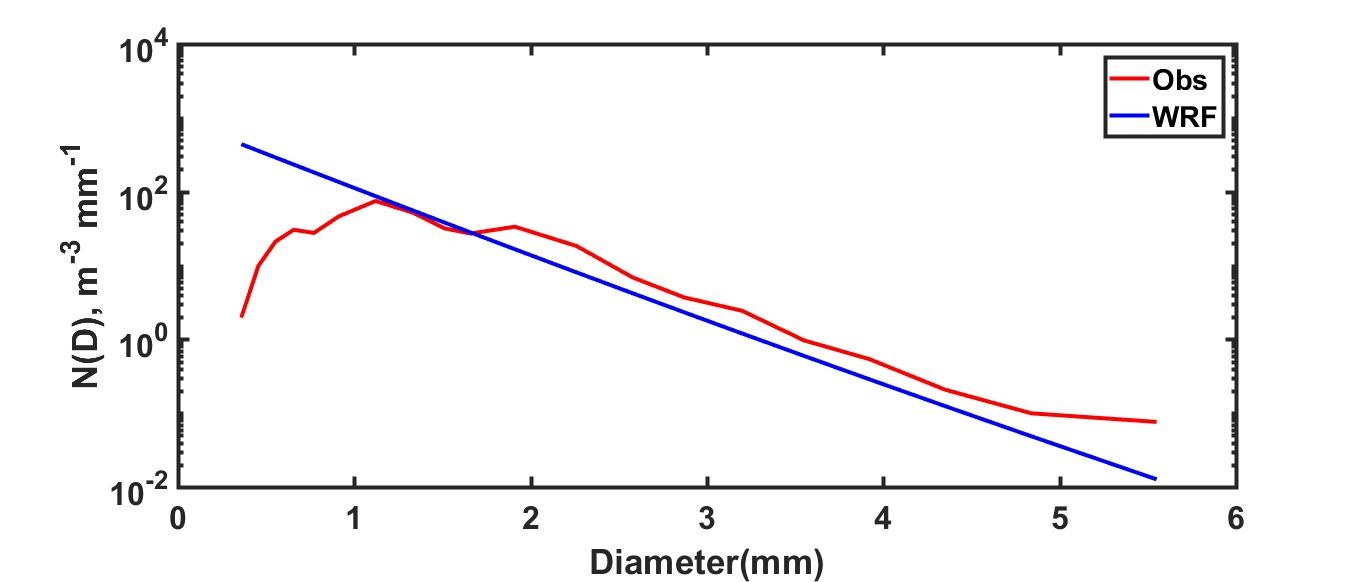}}
\caption{Rain drop size distribution (RDSD) from NWP model (WRF) and in situ observation over High Altitude Cloud Physics Laboratory (HACPL), India (Lon: 73.66 $^0$E and Lat: 17.92 $^0$N). The RDSD observations were made by a Joss Waldvodgel Disdrometer in the diameter range 0.3-5.5 mm.}
\label{fig-1}
\end{figure}

In recent decades weather forecasting has led to significant improvement in the simulation of precipitation in synoptic and mesoscales by Numerical Weather Prediction (NWP) models \cite{Boe14}. But, the problem of errors in the quantitative precipitation forecast (QPF) still appears, which is related to the phase of the diurnal cycle of precipitation over land \cite{Dir12} and underestimation (overestimation) of moderate and heavy (lighter) precipitation \cite{Ken12}. The same factors are also responsible for the skillful predictions of frequency of extreme rainfall events \cite{Gos06}. The overestimation (underestimation) of light (heavy) rain is also reported in Weather Research $\&$ Forecasting (WRF) model \cite{Mud18,Mud22}. A large part of the overestimation (underestimation) of the light (heavy) rainfall intensity may be related to the formation of the raindrop size distribution (RDSD) in the clouds from regional climate models (e.g., WRF) and observation (Figure \ref{fig-1}). The model overestimates (underestimates) smaller (bigger) rain drop sizes (Figure \ref{fig-1}). Many previous studies have already highlighted the underlying problem in coupled climate model like CFSv2 \cite{Haz15, Haz16, Pat13, Saha14}.
\par
\begin{figure}[htbp]
\centerline{\includegraphics[height=16cm,width=12cm]{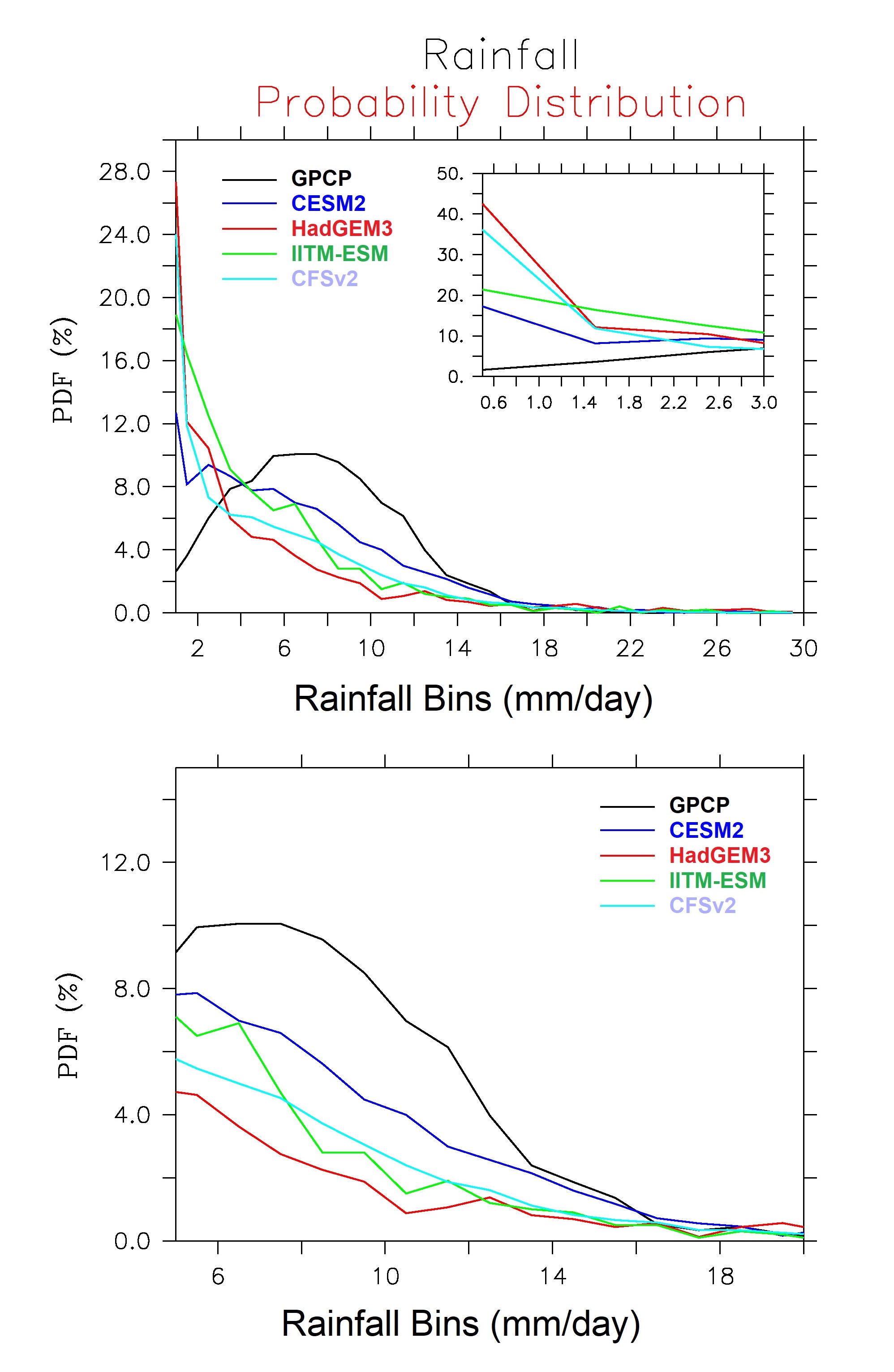}}
\caption{The probability distribution of rainfall (mm/day) over central India (Lon: 74 - 83 $^0$E; Lat: 17 - 27 $^0$N) for different CMIP6 climate models (CESM2, HadGEM3 and IITM-ESM) along with CFSv2. The PDF is also compared with observed rainfall from GPCP.}
\label{fig-2}
\end{figure}

Earlier studies have highlighted that the majority of coupled global climate model participating in the Coupled Model Intercomparison Project (CMIP) is overestimated light rain and underestimated moderate and heavy rain \cite{Sab13,Gos14,GosGos17}. It is important to note that the present generation CMIP6 models also overestimates (underestimates) lighter (moderate and heavy) rainfall (Figure \ref{fig-2}). 
In this regard, the relationship between evaporation-precipitation distribution and its variability over Indian Ocean is also crucial for ISM \cite{Pok12}. \citet{Saha17} have also shown that modification of snow scheme in climate model can improve teleconnection and seasonal mean ISM rainfall. Recently, \citet{Pra22} have demonstrated that the improvement in dirunal cycle can improve the boundary condition between atmosphere and ocean, which finally improve ISMR. The improved ocean initial condition is also important for the climate model in simulating better ISMR \cite{Gag22}. Apart from other physical processes and initial/boundary conditions, the cloud and microphysics are similarly important for the depiction of ISMR \cite{Haz17,Abh17}. It is also known that an adequate cloud parameterization \cite{Phan23,Gana19} and particularly parameterization of ‘cloud microphysics’ is essential for the simulation of ISM and large-scale mesoscale systems \cite{Haz17,Haz20,Mud22,Abh17}. Therefore, understanding and accurately representing physical processes involved in the formation or growth of clouds and rain droplets is important. 
\par
In the early stage of cloud development, after condensation mixing between cloud droplets and atmospheric air governs the diffusional growth of water vapors \cite{PruKle10}. It plays a vital role on the evolution of cloud droplet size distribution (DSD), which influences the rate of formation of warm cloud precipitation in cumulus clouds. After that, cloud to rain water autoconversion and collisional growth are another most important process that controls the formation of warm rain  ~\cite{PruKle10}. Studies have shown that the “autoconversion” initializes the precipitation formation \cite{WuXiDonZha18} and also responsible for the formation of drizzle in stratiform clouds \cite{LiuDauMc04}. However, due to unrealistic representation of threshold behavior in Kessler type autoconversion parameterization, \citet{Sun78} suggested an alternative autoconversion parameterization that explicitly considers cloud liquid water content as the only variable, in contrast with that of \citet{Kes69}. Later, \citet{LiuDauMc04, LiuDauMc06} generalized the Sundqvist type parameterizations.
\par
The humidity distribution within the tropical region is determined by many factors, including the detrainment of vapour and condensed water from convective systems and the large-scale atmospheric circulation. Most of the cloud schemes in climate model are based on the relative humidity (RH) and cloud cover \cite{ZaoCar97,HanPan11}. The cloud can be formed when RH $<$ 100 $\%$, implicitly assuming sub-grid scale variability for total water. Hence, in contrast to cloud feedback, a strong positive water vapour feedback is a robust feature of climate models \cite{Sto01}, being found across models with many different schemes for advection, convection, and condensation of water vapour. The condensation, diffusional growth, and evaporation rate of cloud condensate, where the relative humidity is lower than the critical value, are required \cite{ZaoCar97}. But it is also important to note that the actual PDF (Probability density function; the shape) of total water and its variance (width) over the Indian region are not so far investigated. Therefore, understanding the growth rate of cloud condensates for different clouds (e.g., shallow, less RH; convective, higher RH) using particle-by-particle based small-scale model is essential. Condensation and evaporation rate of cloud condensate depend upon the critical RH \cite{ZaoCar97}, which is discussed in the manuscript of Part-I by \citet{Bhow23a}.
\par
Therefore, to improve the numerical model, understanding the variation of the diffusional growth rate coefficient, relative dispersion are important, which can be obtained from direct numerical simulation (DNS) using own observations. The diffusional growth of cloud droplets from small-scale parcel model has already been used into NWP model (e.g., MM5 and WRF) as reported by \citet{CheWanChe07,CheWanChe10} and \citet{ThomFieRas08}. \citet{WuYu10} have shown that any 'tuning' parameter needs to be modified based on the resolution of climate models. \citet{MarJohSpi94} have developed parameterization for climate from a single point observational measurements. Therefore, any new parameteriation and 'tuning' parameters obtained from small scale model (e.g., DNS) should be added value in NWP model community and important for climate modelers to do sensitivity experiments with range of values suggested by DNS experiments. In this present endeavor (Part II), it will be demonstrated the usefulness of these parameters and new parameterizations through climate model sensitivity experiments with high resolution ($\sim$ 1 km) WRF real simulation followed by coarse resolution ($\sim$ 38 km) climate model (CFSv2) for the detailed ISMR study. We have simulated with dispersion based “autoconversion” in climate forecast system (CFS) and modified the diffusional growth rate of cloud droplets. 
\par
The present paper is organized as follows. The parameterization and numerical model experiments with WRF and CFSv2 is discussed in section 2. The results of PDF (Probability density function) from sensitivity studies are detailed in Section 3. Conclusions are summarized in Section 4.
\section{The parameterization and model experiments:}
The warm rain (autoconversion) is a major process, which affects the raindrop budget and plays a more important role in ISM precipitation \cite{Haz16}. It should be noted that this autoconversion parameterization scheme is very commonly used in general circulation models \cite{Loh07,MorGet08}. Some attempts have been made for the tunning of autoconversion coefficients in climate models for better simulation of ISM \cite{Haz17, Gana19, Dutt21}. In the coupled climate model (CFSv2), the cloud water to rain water autoconversion is followed by \citet{Sun89}, which is primarily proposed by \citet{Sun78}. The expression for the Sundqvist-type autoconversion rate is:
\begin{equation}
P_{raut} = c_0 ~q_l \left[1-exp\left(-\left(\frac{q_l}{q_{l_{crit}}}\right)^2\right)\right]
\end{equation}
where, $q_l$ is the cloud water content and $q_{l_{crit}}$ is the threshold (or critical) cloud liquid water content. $c_0$ is an empirical constants. 
\par
Kessler-type autoconversion \cite{Kes69} is very simple and commonly used scheme, which assumed that the precipitation rate is directly proportional to the cloud water content. Later, \citet{Sun78} proposed an alternative expression for the autoconversion rate. \citet{LiuDauMc04,LiuDauMc06} proposed the relative dispersion based autoconversion scheme, which assumes that the autoconversion rate is related to the cloud water content, droplet number concentration, and relative dispersion of cloud droplets. \citet{LiuDauMc04} generalized the Sundqvis-type parameterizations. The relative dispersion based autoconversion is more realistic than that of conventional Kessler or Sundqvist type. It is also important to note that small-scale Lagrangian particle based numerical simulation can provide the value of relative dispersion to formulate Liu and Daum autoconversion rate. Therefore, in this present study, cloud water to rain water autoconversion of \citet{Sun89} is further modified using small-scale numerical model based on Cloud Aerosol Interaction and Precipitation Enhancement EXperiment (CAIPEEX) observation \cite{Bhow23a}. The threshold (critical) cloud liquid water content is obtained from \cite{RotLiu05}.
\begin{equation}
    q{{_l}_{crit}} = \frac{4}{3} \pi \rho_w \beta^{-3} r_{crit}^3 N_c
\end{equation}
\begin{equation}
    \beta = \left[ \frac{(1+3\epsilon^2)(1+4\epsilon^2)(1+5\epsilon^2)}{(1+\epsilon^2)(1+2\epsilon^2)}\right]^{1/6}
\end{equation}
where $N_c$ is the number of cloud droplets collected from CAIPEEX observation over Indian subcontinent. $\epsilon$  is relative dispersion obtained from DNS (as discussed in the manuscript Part I of \citet{Bhow23a} and verified with air borne observation and other previous study). 
$r_{crit}$ is a critical (threshold) cloud droplet size for the parameterization of cloud water to rain water autoconversion as shown by several studies \cite{RasKri98,Rot00}. In this study, the value of $r_{crit}$ is obtained from CAIPEEX observation and DNS simulation \cite{Bhow23a}.

The mass growth rate of cloud droplets \cite{PruKle10,LamVer11} can be presented as, 
\begin{equation}
    \frac{dq_l}{dt} = 4 \pi r_d M_w D_v (n_s - n_{eq})
\end{equation}
Where,  $M_w$  is the molar mass of water, $D_v$ is the vapour diffusivity, $n_s$ and $n_{eq}$ are the vapor profile in steady-state and equilibrium conditions, respectively. It is important to note that the gradient of growth rate is directly proportional to excess vapor $(n_s- n_{eq})$ and indirectly proportional to radius of cloud droplets. The surface-boundary condition ($n_{eq}$) is highly sensitive to the temperature of the particle. The temperature can also effect the equilibrium vapour pressure (The Clausius-Clapeyron equation). The cloud droplets are neither pure or flat, however, we must modify the equilibrium vapour pressure to account the solute effect and the curvature effect as provided by K$\ddot{o}$hler theory \cite{LamVer11}. Therefore, the expression of diffusional growth rate of cloud droplets is,
\begin{equation}
    \frac{dq_l}{dt}=4 \pi r_{drop}^2 \rho_l \left(\frac{dr_{drop}}{dt}\right)
\end{equation}
Where, $\frac{dr_{drop}}{dt}$  is the growth rate cloud droplets, the formulation is available in (equation 5 of \cite{Bhow23a}). The new mean growth rate coefficient for convective clouds is 1.5 $\times$ 10$^{-5}$ m/s, which can be obtained from small scale model simulation \cite{Bhow23a}.  

\par
The climate forecast system version 2 (CFSv2) \cite{Saha14} model has been selected as a base model for the future development of a reliable seasonal prediction system of the ISMR under the Monsoon Mission Project (http://www.tropmet.res.in/monsoon/) of the Ministry of Earth Sciences, Government of India \cite{Rao19}. The new SAS (nSAS) \cite{HanPan11} convective parameterization scheme is believed to make cumulus convection stronger and deeper, and to deplete more instability in the atmospheric column. Two sensitivity experiments in CFSv2 are carried out such as, (i) control (CTL) based on Sundqvist-type parameterization and (ii) dispersion based Liu and Daum autoconversion rate with nSAS convective parameterization scheme. For both simulations, other physical processes (e.g., land-surface, radiation, etc.) are kept the same. The details of the model are available in \citet{Saha14}. In each experiment, the CFSv2 has been initialized by same initial conditions and model is integrated for 15 years. Climatological means of required fields are prepared by considering last 10 years.
\section{Results from climate model sensitivity experiments}
\begin{figure}[htbp]
\centerline{\includegraphics[height=16cm,width=12cm]{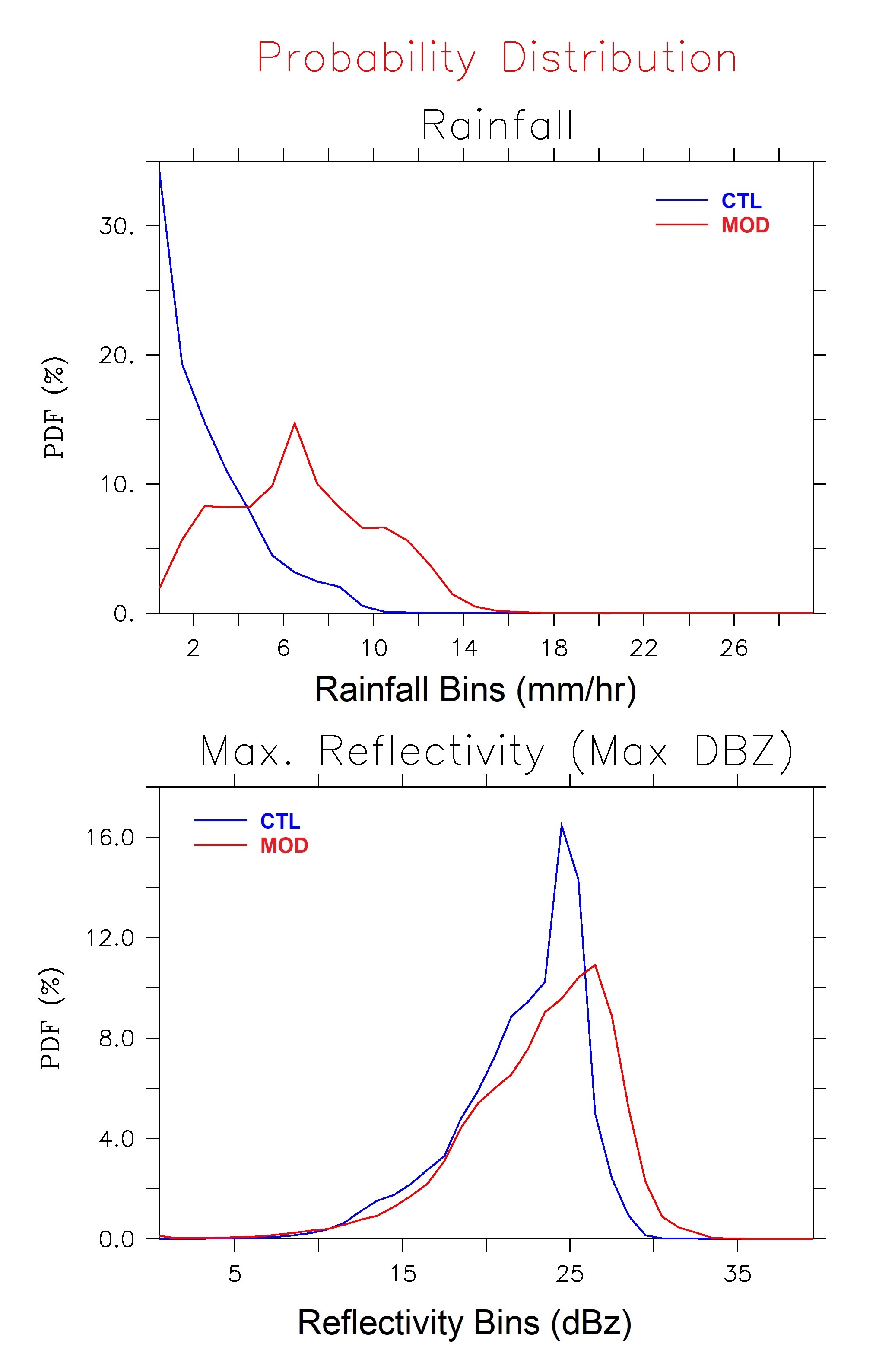}}
\caption{TThe probability distribution of rainfall (mm/hr) and maximum reflectivity (Max. DBZ) for the convective cloud over Maharashtra, India (Lon: 73 - 78$^0$E; Lat: 14 - 20$^0$N) for WRF (control, CTL and modified, MOD).}.
\label{fig-3}
\end{figure} 
\subsection{Probability distribution and mean of ISM rainfall}
The low intensity rainfall events which represent the crest of probability distribution function (PDF) and inversely moderate or high intense rainfall that signify the tail of PDF are certainly important for the daily variance of ISM rainfall \cite{GosGos17}. To demonstrate the role of cloud microphysics parameterization (diffusional growth rate coefficient and autoconversion) obtained from DNS using air borne observation as initial condition and to evaluate the model’s ability to simulate the precipitation distribution, we have examined the PDF for two sensitivity experiments (i.e., control, CTL and modified, MOD). Firstly, the high resolution (1 km horizontal resolution) WRF for 3-day simulation for convective cloud during Indian summer monsoon (ISM) is used for preliminary understanding of rainfall and maximum reflectivity PDF (Figure \ref{fig-3}), which can provide confidence to introduce then into the climate model. Interestingly, the frequency of lighter (moderate) rain decreases (increases) in MOD experiment as compared to CTL (Figure \ref{fig-3}). The PDF of maximum reflectivity (at surface), which is proportional to size of hydrometeors (i.e., rain drops here) is also examined (Figure \ref{fig-3}) to validate the rainfall PDF. The frequency of reflectivity in lesser (higher) bin decreases (increases) and the maximum reflectivity slightly shifts towards higher bin in case of MOD experiment than CTL (Figure \ref{fig-3}). The PDF results of rainfall and reflectivity from 3-day simulation of high resolution WRF experiments demonstrate that physically based modified diffusional growth rate and autoconversion are crucial in modifying crest and tail of PDF. 
\begin{figure}[htbp]
\centerline{\includegraphics[height=12cm,width=14cm]{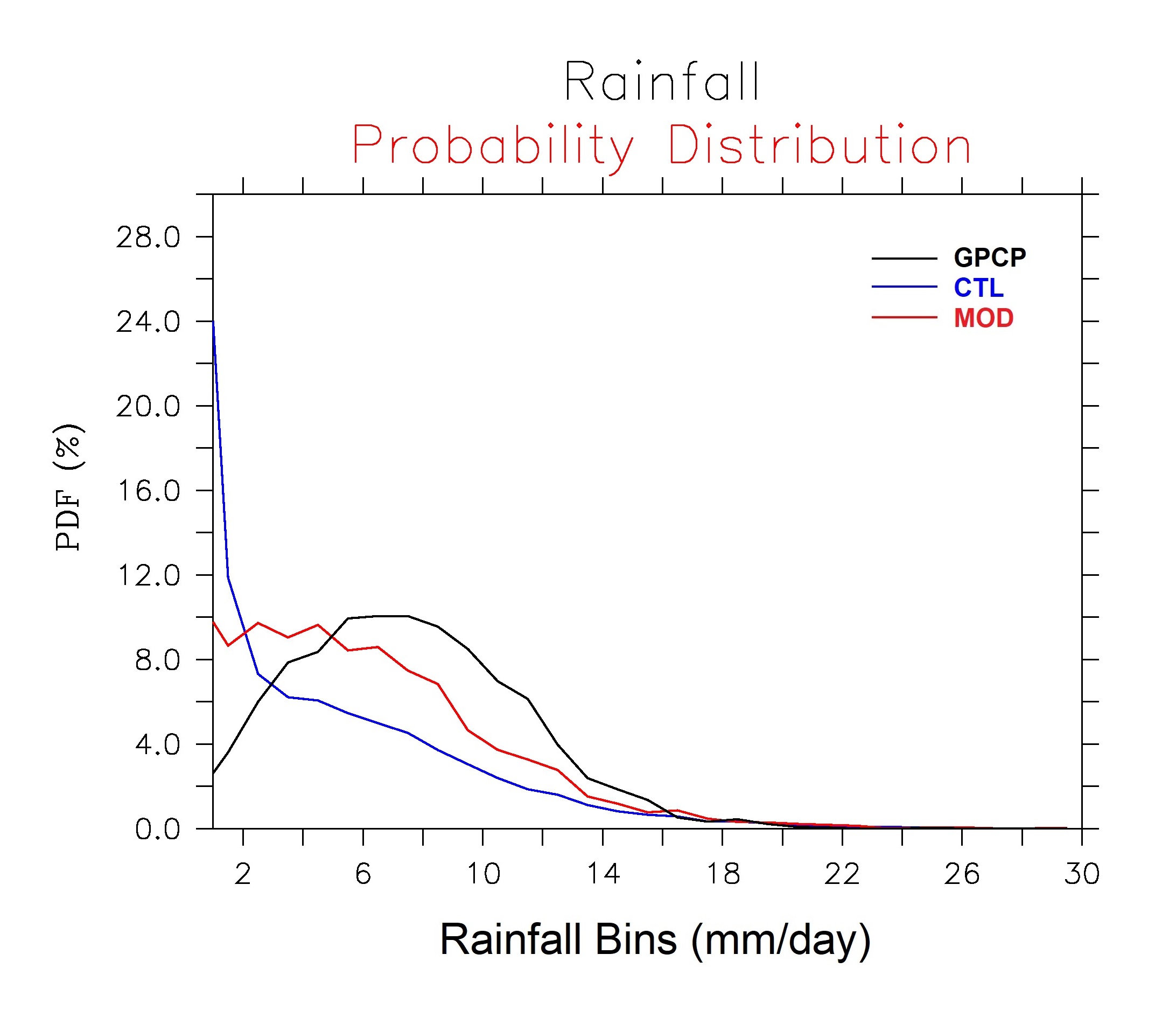}}
\caption{The probability distribution of rainfall (mm/day) over central India (Lon: 74 - 83 $^0$E; Lat: 17 - 27 $^0$N) for CFSv2 (control, CTL and modified, MOD) and observation (GPCP).}.
\label{fig-4}
\end{figure} 

Therefore, as our focus is to see the improvement in ISM rainfall, we have conducted experiments with coupled climate model (CFSv2) with control (original parameterization) and modified parameterization (new diffusional growth rate coefficient and autoconversion). In this case, the PDF of more variables (cloud, convection and thermodynamic) are also analyzed from coupled climate model only. The observed and two models using coupled climate model (CFSv2) PDF of rainfall are plotted in figure \ref{fig-4}. It can be noted that the control CFSv2 overestimates the crest (the lighter rain events) and underestimate the tail (the moderate and heavy rainfall events) of the rainfall distribution for central India region (Figure \ref{fig-4}). The problem of overestimation of low rain events and underestimation of moderate and heavy rain events appears to be generic and fundamental problem of all CMIP6 climate models (Figure \ref{fig-2}). The physically based modified parameterization in CFSv2 (MOD experiment) shows the improvement in the overestimation ($\sim$ 14 $\%$) of lighter rainfall event and underestimation of moderate and heavy rainfall events (Figure \ref{fig-4}). There is also significant improvement of rainfall in 5-16 mm/day rain bins for MOD experiment (Figure \ref{fig-4}). The underestimation in that rain category (5-16 mm/day rain bins) reduced distinctly as compared to control experiment. But still there is a gap relative to observation, which might be taken care by introducing other important processes responses for the growth of larger droplets. Earlier studies \cite{Haz13,KumHaz14} demonstrated that there are strong interactions between aerosols, dynamics, cloud microphysics, and other physical processes in NWP model, which also play a role in the formation rain in different categories. 

\begin{figure}[htbp]
\begin{center}
{\includegraphics[height=16cm,width=20cm]{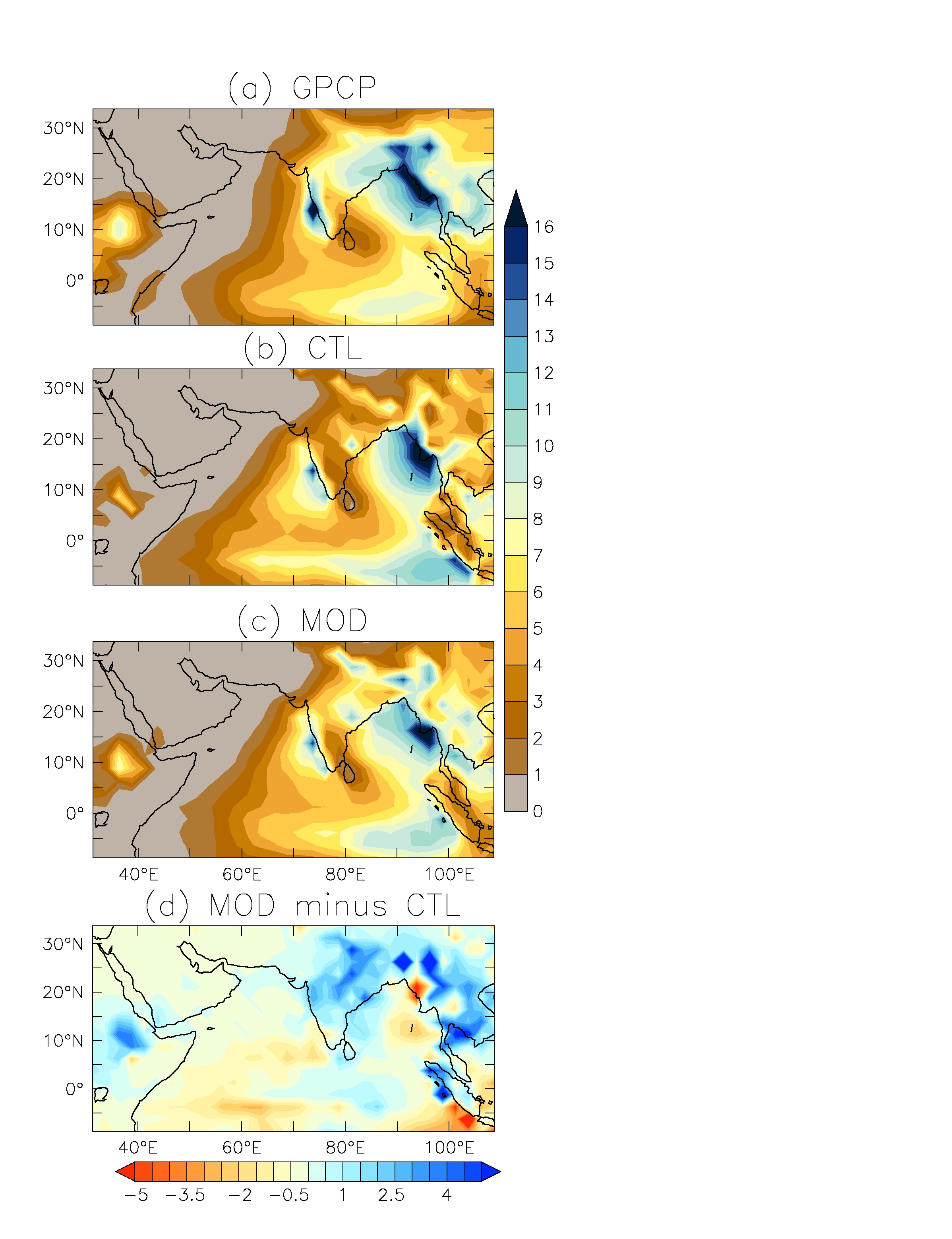}}
\end{center}
\caption{The JJAS (June to September) mean rainfall (mm/day) over extended ISM region for observation (GPCP) and CFSv2 (control(CTL), modified (MOD) and difference (MOD - CTL)}.
\label{fig-5}
\end{figure}
\begin{figure}[htbp]
\begin{center}
{\includegraphics[height=16cm,width=20cm]{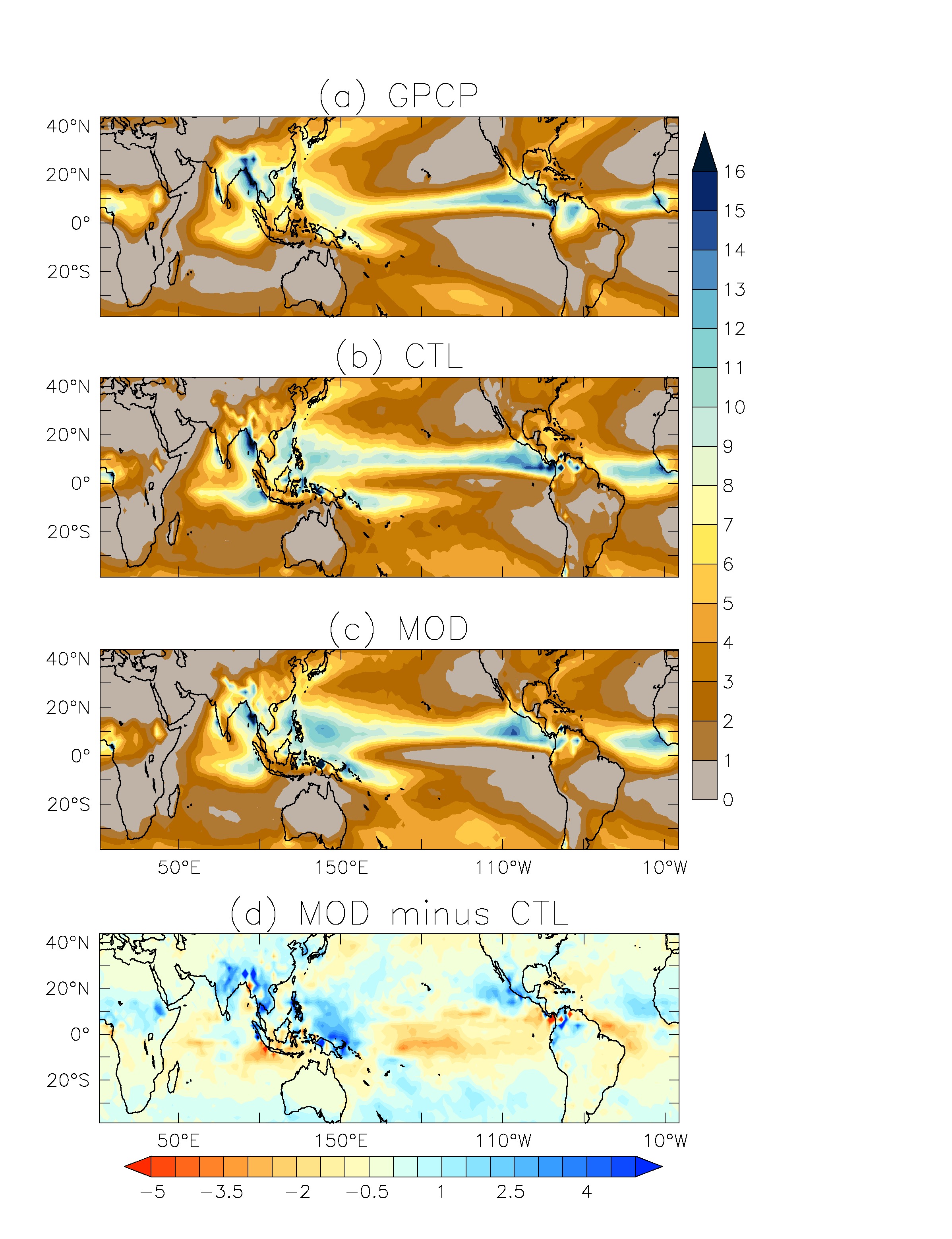}}
\end{center}
\caption{The JJAS (June to September) mean rainfall (mm/day) over global region for observation (GPCP) and CFSv2 (control(CTL), modified (MOD) and difference (MOD - CTL)}.
\label{fig-5b}
\end{figure}

 
The seasonal (June to September, JJAS) mean rainfall from observation (GPCP) (Figure \ref{fig-5}a) and two sensitivity experiments with CFSv2 (control (CTL) and modified (MOD)) (Figure \ref{fig-5}b and Figure \ref{fig-5}c) and their differences (MOD-CTL) (Figure \ref{fig-5}d)  over Extended Indian Monsoon Region (EIMR) region (i.e over Western Ghats, central India along with north-east region) are plotted in Figure \ref{fig-5}. It is clearly indicates the improvement in spatial pattern of JJAS mean rainfall over Extended Indian Monsoon Region (EIMR) region (Figure \ref{fig-5}). Similarly, the global JJAS mean rainfall patterns from observation (GPCP) (Figure \ref{fig-5b}a) and two sensitivity experiments with CFSv2 (control(CTL) and modified (MOD)) (Figure \ref{fig-5b}b and Figure \ref{fig-5b}c) and there differences (MOD-CTL) (Figure \ref{fig-5b}d) are shown in Figure \ref{fig-5b}. A clear improvement in the JJAS mean rainfall particularly over central India and its adjacent regions can be observed from both the figures (Figure \ref{fig-5} and Figure \ref{fig-5b}). The persistent spatial dry rainfall biases over Indian land mass improved significantly (Figure \ref{fig-5}d).  

\begin{figure}[htbp]
\centerline{\includegraphics[height=7cm,width=14cm]{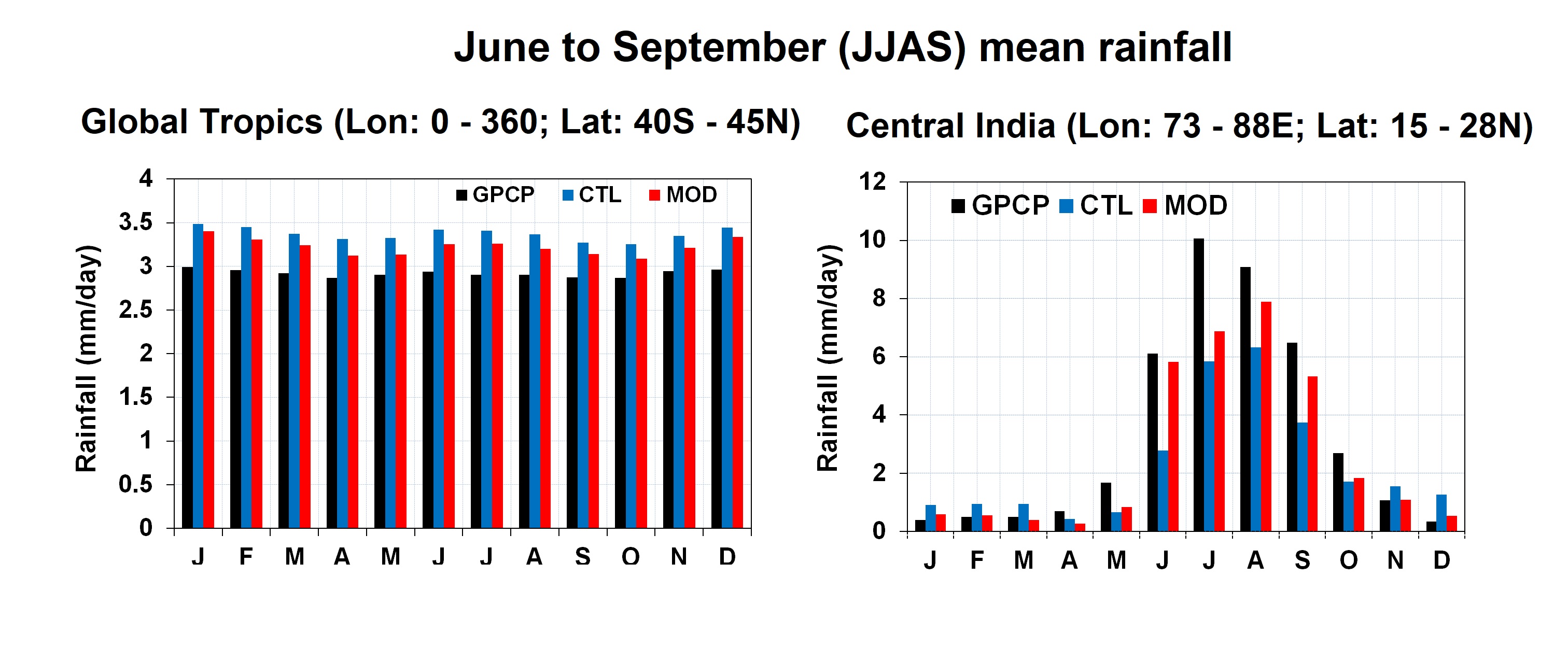}}
\caption{The annual cycle of rainfall (mm/day) over global tropics (Lon: 0 - 360; Lat: 40 $^0$S - 40 $^0$N) and central India (Lon: 74 - 83 $^0$E; Lat: 17 - 27 $^0$N) for CFSv2 (control, CTL and modified, MOD) and observation (GPCP).}.
\label{fig-6}
\end{figure}

The Annual cycle of rainfall over global tropics and central India are also shown (Figure \ref{fig-6}). The results demonstrate that MOD simulation reduce the wet (dry) rainfall bias over global tropics (central India) (Figure \ref{fig-6}). The DNS findings already show how the Sundqvist type and Liu-Daum type of autoconversion vary with cloud liquid water content (see Figure 6 of Part-I \cite{Bhow23a} of this manuscript). This clearly indicates that drizzle-like rainfall can be modified and moderate-like rainfall can also be improved due to better cloud-to-rain mass expansion.
Now, the reduction of biases (mean and PDF) must be linked with the thermodynamical and cloud variables. To understand the physically linked or consistent development in rainfall production, we have also plotted the PDF and mean of specific humidity, cloud condensate (liquid water content) and outgoing long-wave radiation (OLR).
%
\subsection{Probability distribution of thermodynamical and cloud variables:}
\begin{figure}[htbp]
\centerline{\includegraphics[height=10cm,width=12cm]{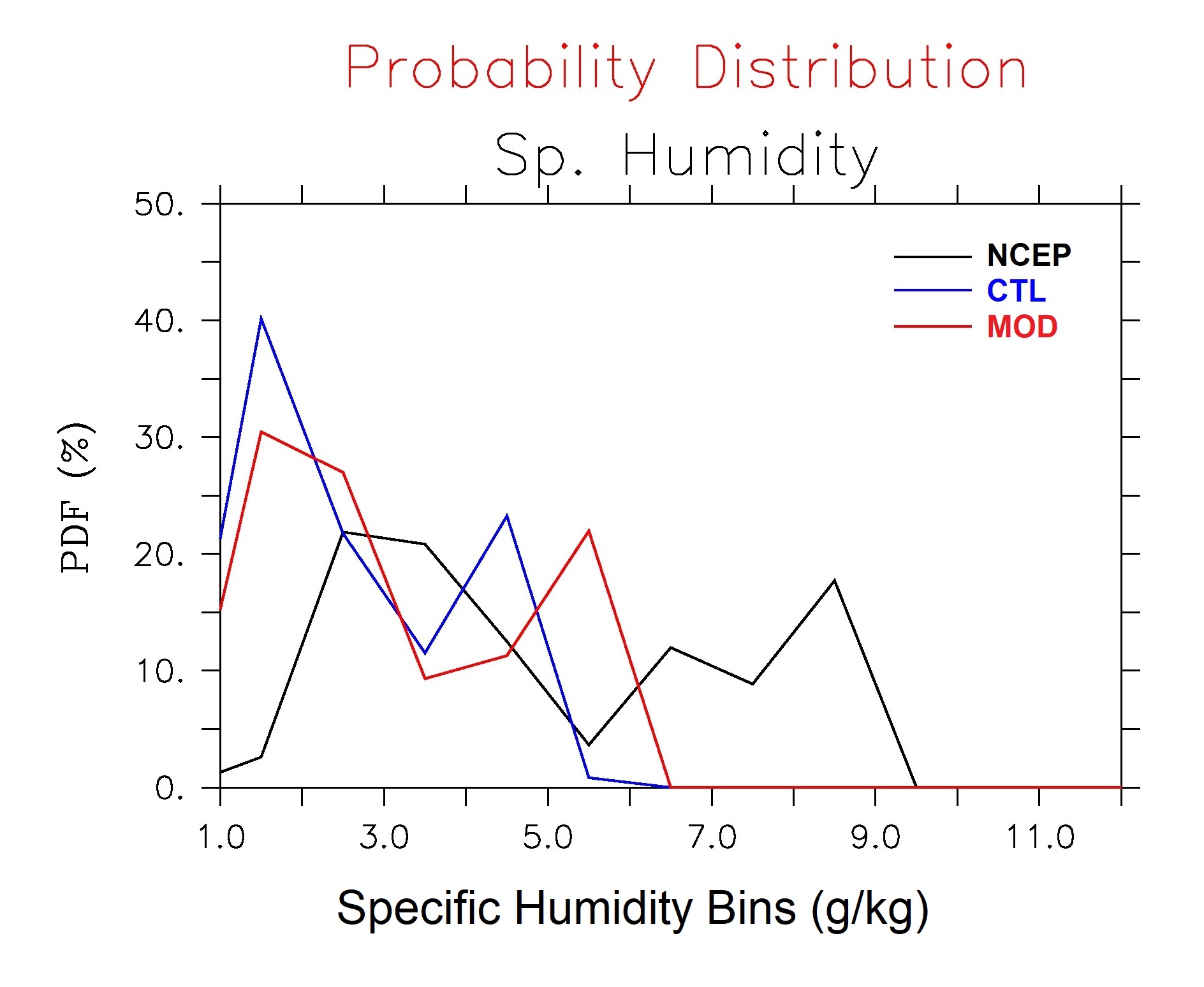}}
\caption{The probability distribution of specific humidity (g/kg) over central India (Lon: 74 - 83 $^0$E; Lat: 17 - 27 $^0$N) for CFSv2 (control, CTL and modified, MOD) and NCEP reanalysis.}.
\label{fig-7}
\end{figure}
\begin{figure}[htbp]
\centerline{\includegraphics[height=9cm,width=13cm]{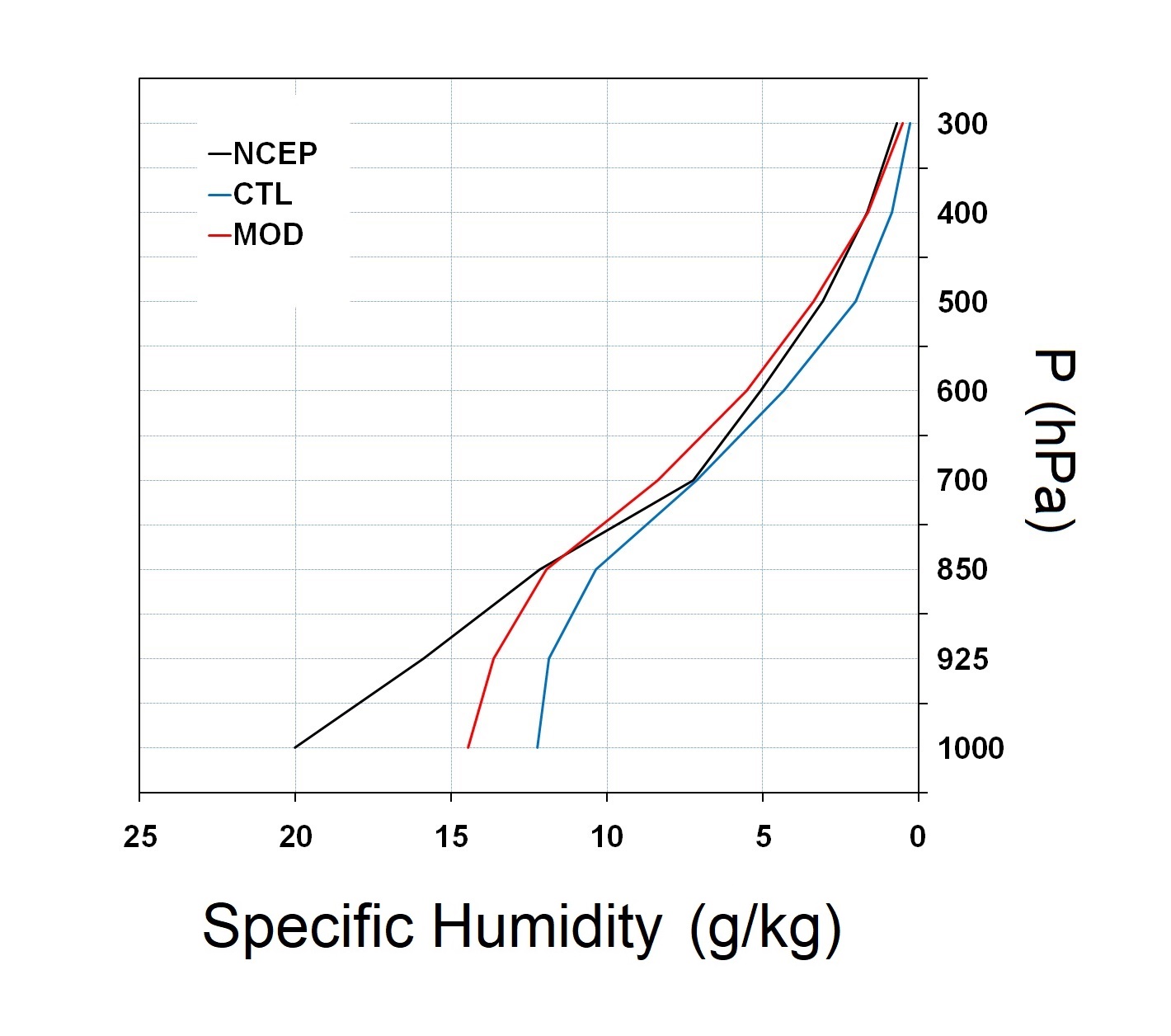}}
\caption{The vertical profile of specific humidity (g/kg) averaged over central India (Lon: 74 - 83 $^0$E; Lat: 17 - 27 $^0$N) for CFSv2 (control, CTL and modified, MOD) and NCEP reanalysis.}
\label{fig-8}
\end{figure}

 
Figure \ref{fig-7} and Figure \ref{fig-8} depict the PDF and mean of specific humidity respectively from CTL and MOD experiments, which has been compared with NCEP. The vertical profile of area averaged (over central India, CI) JJAS mean specific humidity bias reduced significantly in MOD as compared to CTL (Figure \ref{fig-8}). The specific humidity in models overestimate the crest and underestimate in the tail as compared to NCEP (Figure \ref{fig-7}). The MOD simulation is able to reduce overestimation (underestimation) in the crest (middle) as compared to CTL, which is connected with overestimation (underestimation) of lighter (moderate) rainfall (Figure \ref{fig-4}). It is noted that still models have an issue to simulate specific humidity in the tail end, which need to be focused in future. The modified growth rate coefficient and autoconversion are responsible for the distribution and conversion of cloud and rain droplets as seen in the DNS experiments \cite{Bhow23a} and also noticed in the maximum reflectivity plots from WRF simulation (Figure \ref{fig-3}). It is known that microphysical processes can also modify latent heat release through condensation and evaporation and further provide feedback to thermodynamics and dynamics \cite{Haz13, Haz17}.

\begin{figure}[htbp]
\centerline{\includegraphics[height=14cm,width=14cm]{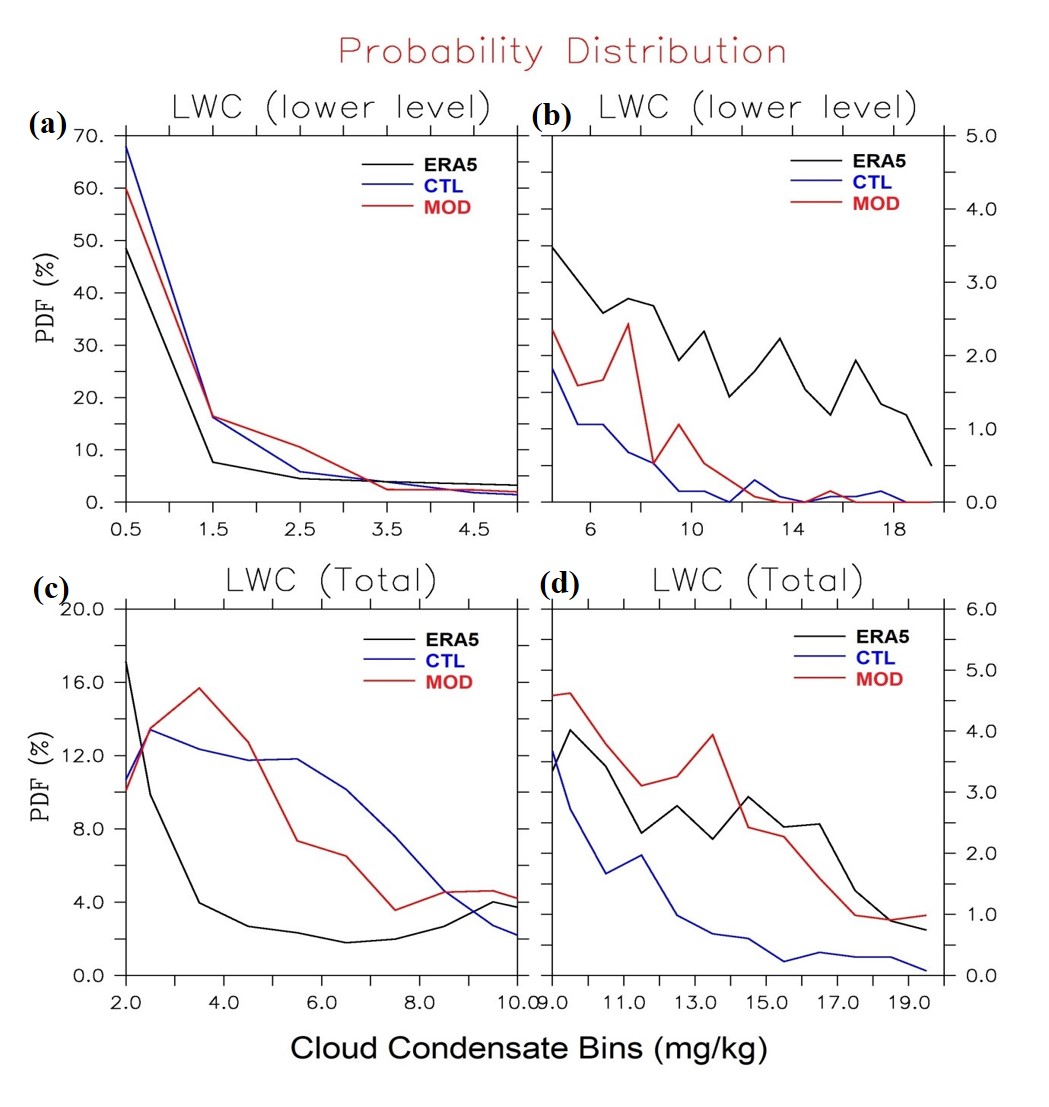}}
\caption{The probability distribution of cloud condensate (liquid water content, LWC, mg/kg) over central India (Lon: 74 - 83 $^0$E; Lat: 17 - 27 $^0$N) for CFSv2 (control, CTL and modified, MOD) and ERA5 reanalysis.}
\label{fig-9}
\end{figure}

The probability distribution of cloud condensate (liquid water content (LWC) mg/kg) over central India for models (CTL and MOD) and ERA5 reanalysis are shown in Figure \ref{fig-9}. The PDF of low level LWC (up to 850 hPa) and total vertical averaged LWC have been depicted here (Figure \ref{fig-9}). The low level cloud condensate (LWC) overestimates ($\sim$ 20$\%$ more than ERA5), which is basically responsible of lighter (drizzle like) rain in model (CTL). The modified parameterization (MOD) is able to reduce the overestimation ($\sim$ 10$\%$) as compared to CTL (Figure \ref{fig-9}a) due to better diffusional growth rate and cloud to rain water autoconversion. Similarly, the underestimation in the middle and tail part of LWC is also linked with the underestimation of moderate and heavy rainfall. The MOD experiment is able to reduce underestimation in middle part of low level LWC bins (Figure \ref{fig-9}b). Similarly, the overestimation (underestimation) of total LWC (all levels) in the PDF crest (tail) has developed in MOD as compared to CTL (Figure \ref{fig-9}c-\ref{fig-9}d). 
\subsection{Probability distribution of outgoing longwave radiation (OLR):}
\begin{figure}[htbp]
\centerline{\includegraphics[height=12cm,width=15cm]{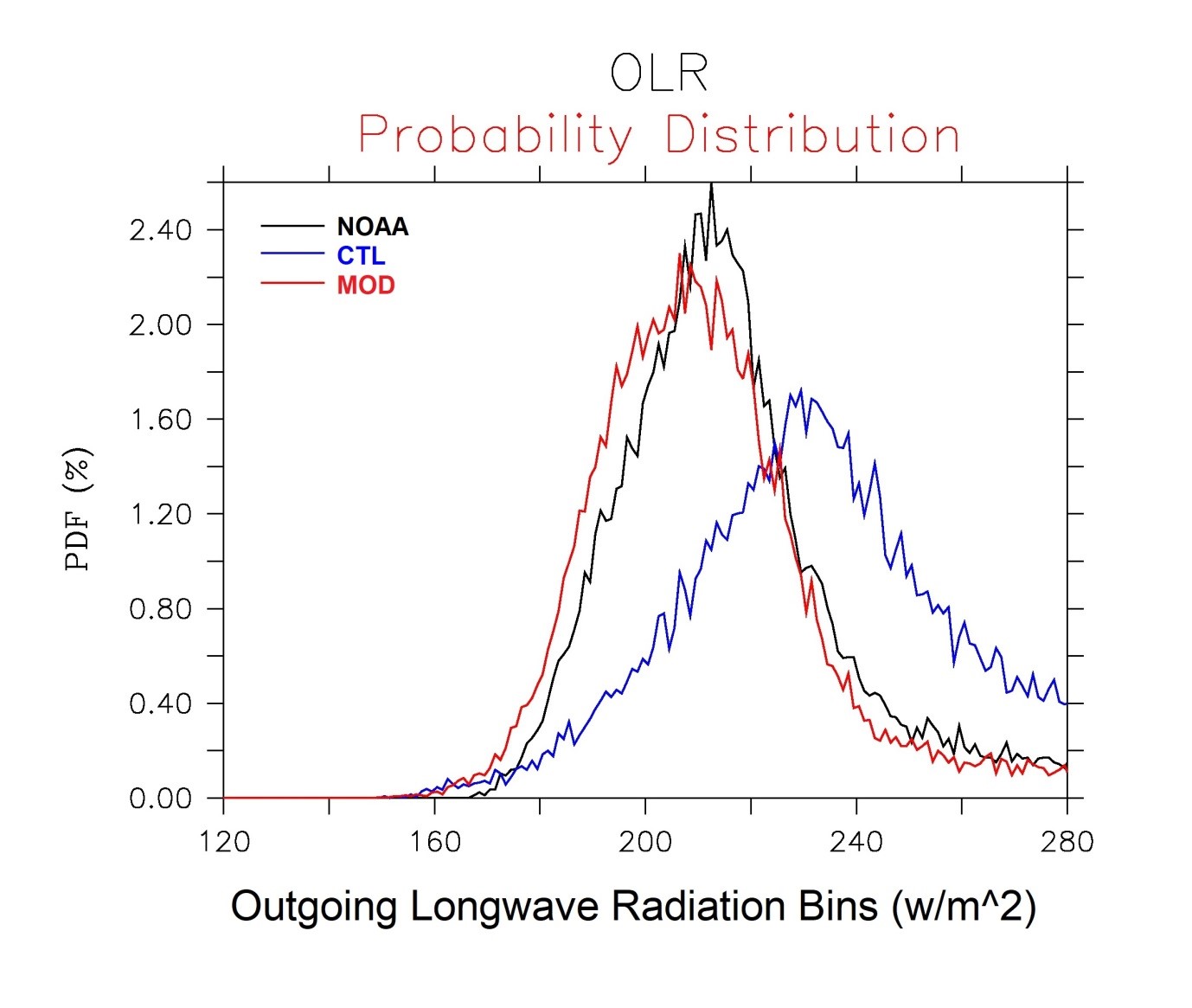}}
\caption{The probability distribution of outgoing longwave radiation (OLR, watt/m2) over bigger ISM region (Extended Indian Monsoon Rainfall, EIMR region, Lon: 70 - 100 $^0$E; Lat: 10 - 30 $^0$N) for CFSv2 (control, CTL and modified, MOD) and NOAA.}
\label{fig-10}
\end{figure}

There are several studies \cite{Kri89,Mur80,Chau10,Praver85}, which demosntrate that the outgoing longwave radiation (OLR) can be considered as a proxy of convection and is essential for monsoon circulations. It is also known that clouds play a seminal role in regulating OLR \cite{Haz17,Cha19,Dutt21}. Here, we have tried to see whether the new diffusional growth rate and dispersion based Liu-Daum type “autoconversion” could modulate the PDF of OLR. The OLR data were taken from National Oceanic and Atmospheric Administration (NOAA) \cite{LieSmi96}. The probability distribution of OLR over larger ISM region (Extended Indian Monsoon Rainfall, EIMR region, Lon: 70 – 100 E; Lat: 10 – 30 N) for CFSv2 (CTL and MOD) are plotted in Figure \ref{fig-10}. The OLR in control version of model highly overestimate the tail (higher OLR bins) and underestimate the crest (lower OLR bin) (Figure \ref{fig-10}). The higher (lower) OLR signify the deep (shallow) convection \cite{Kri89,Mur80}. Therefore, too much shallow convection in CTL model (Figure \ref{fig-10}) is responsible for the ample to lighter rain (Figure \ref{fig-4}), which is a generic problem of all latest generation CMIP6 model also (Figure \ref{fig-2}). In this regard, the physically based microphysical parameterization guided by DNS using in situ observation can shed light on obtaining realistic probability distribution of OLR. The PDF of OLR in MOD simulation significantly improved as compared to CTL (Figure \ref{fig-10}). The frequency of deep convective clouds (low OLR bin) increased in MOD experiment, which is close to observation (Figure \ref{fig-10}). 
\section{Conclusions and Discussions:}
The persistence of the dry bias in simulating ISM rainfall and overestimation of light rain events while underestimate the tail of the PDF (moderate and heavy rain events) in climate models highlights an unresolved problem. In this context, can small-scale cloud model and climate model work better together for improved PDF? The small-scale based modified microphysical parameterizations have shown the improvement in probability distribution of rainfall, OLR, cloud condensate and specific humidity in coupled climate model (CFSv2). We have already examined the diffusional growth rate coefficients, relative dispersion and comparison of two types of ‘autoconversion’ (i.e., Sundqvist type and Liu-Daum type) in Lagrangian particle-by-particle based numerical simulations in the Part-I \cite{Bhow23a} of this manuscript. Recently, \citet{Mor20} have published an insightful work on the confronting the challenge of cloud, microphysics and precipitation to highlights the pathways for modeling community. They mentioned that the Lagrangian particle‐based method, which is used in direct numerical simulation, has gained power as it considered cloud and precipitation particles (called“super‐particles”). Many recent studies \cite{GraWang13,Grab19,Mor20,Chen20} have proposed that to accelerate improvements in microphysics schemes in weather and climate model, process studies are crucial using Lagrangian particle‐based schemes. Here, in this present study, we have demonstrated that the modified parameterization of cloud to rain water autoconversion rate and new growth coefficient in climate model led to a more realistic PDF of precipitation and other cloud variables.
\par
The significant differences in the Sundqvist-type and dispersion based Liu-Daum type ‘autoconversion’ rates in Lagrangian particle‐based schemes motivate us to assess their effect in large-scale climate model. The high resolution (1 km resolution) climate model (WRF) yields the preliminary confidence where it clearly reveal the changes in maximum reflectivity at the surface (related to size distribution of cloud and rain hydrometeors) and finally in PDF of rainfall. The results highlight the importance of model development by incorporating new coefficient of diffusional growth process and cloud droplets spectral dispersion over Indian summer monsoon clouds. Therefore, we may conclude that the biases in the PDF of rainfall, which is a generic problem of almost all climate models can be reduced through physically based microphysical parameterizations. 
\par
\begin{figure}[htbp]
\centerline{\includegraphics[height=14cm,width=12cm]{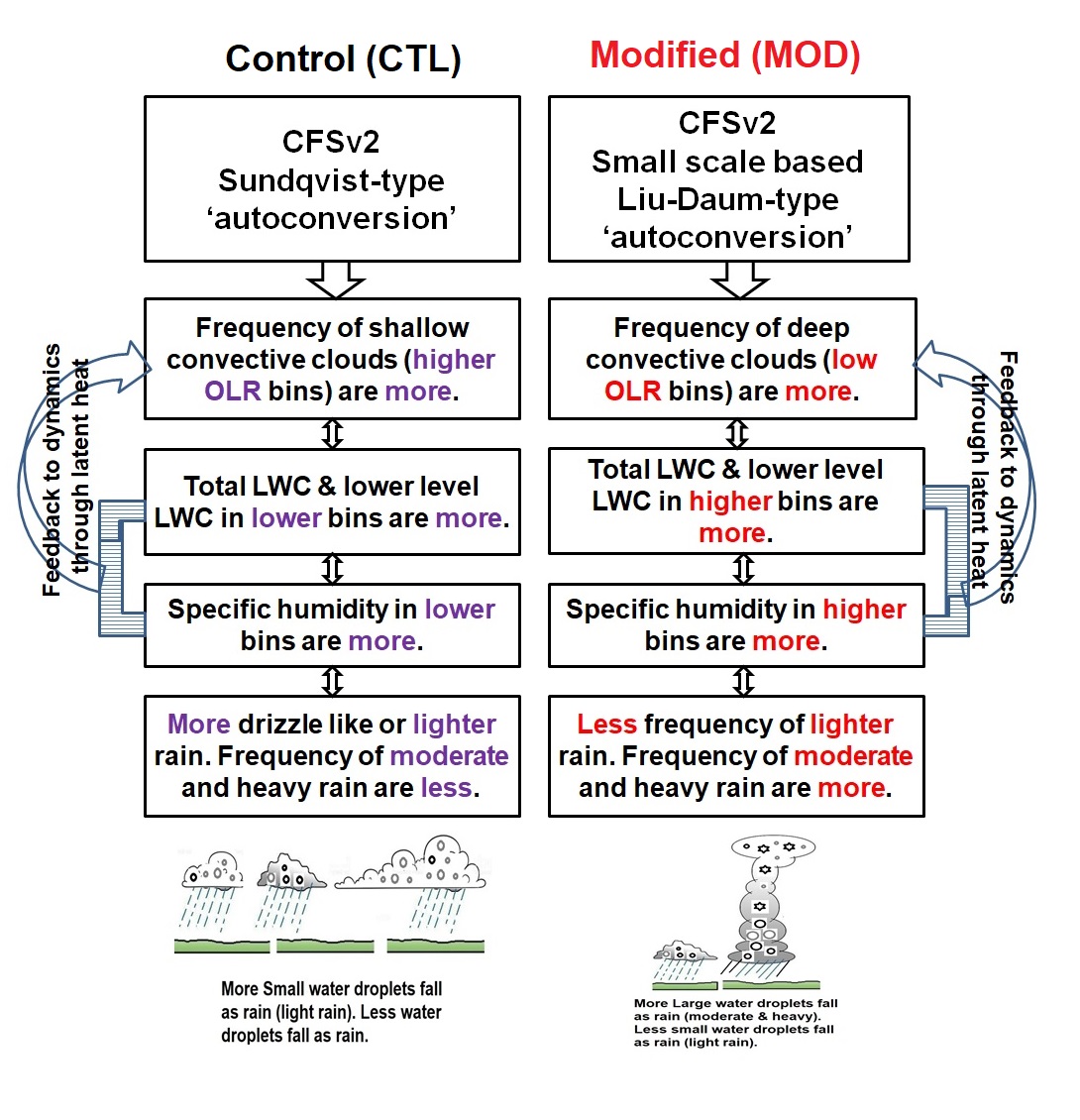}}
\caption{The Schematic diagram illustrating possible mechanism for the betterment of probability distribution of Indian Monsoon Rainfall in CFSv2 (control, CTL and modified, MOD).}
\label{fig-11}
\end{figure}

The development of ISM rainfall (mean and PDF) is also well connected with the PDF of thermodynamical (specific humidity), cloud (LWC) and indicator of convection (OLR). The modified version able to reduce biases as compared to control version. As compared to observation, the modified version's simulation of ISM produces more deep convective clouds, which is more realistic. In contrast, control version simulates more shallow clouds. In a nutshell, to explain the mechanism for two sensitivity experiments using CFSv2, we have presented a schematic diagram in Figure \ref{fig-11}. Proper representation of cloud microphysics parameterization (e.g., autoconversion and diffusional growth rates) cloud condensate, specific humidity and outgoing longwave radiation in MOD experiment leads to improve PDF of rainfall. Thus, the proper representation of autoconversion in microphysical scheme for the global coupled model is important not only because of microphysics but also due to the convection-radiation and dynamical feedbacks (as seen in OLR) (Figure \ref{fig-11}).  \citet{GosGos17} have shown that in CMIP5 model the PDF of rainfall is origin of biases in the simulation of rainfall variance. 
\par
The variance, propagation, spectra and teleconnection are not attempted in this study, which will be done in separate studies. Still there are biases in the tail of all PDFs (rainfall, specific humidity and LWC), which can be targeted by incorporating other important microphysical processes (e.g., collision-coalescence, riming, freezing etc.). This study is purely based on a “physical” approach combined with air borne observations over Indian subcontinent and enables ad hoc “tuning” of several parameters based on the resolution and dynamical core.   
 \newpage
\section{Acknowledgements}
The IITM Pune is funded by the Ministry of Earth Science (MoES), Government of India. Authors thankful to Director, IITM for providing support and encouragement to carry out this research work. The authors acknowledge the use of high-performance computational resources at IITM, particularly "Aaditya" and "Pratyush" HPC system without which this work would not have been possible. 

\begin{thebibliography}{3}
\providecommand{\natexlab}[1]{#1}
\providecommand{\url}[1]{\texttt{#1}}
\expandafter\ifx\csname urlstyle\endcsname\relax
  \providecommand{\doi}[1]{doi: #1}\else
  \providecommand{\doi}{doi: \begingroup \urlstyle{rm}\Url}\fi

\bibitem[Berge(1957)]{Berge57}
Claude Berge.
\newblock Two theorems in graph theory.
\newblock \emph{Proceedings of the National Academy of Sciences}, 43\penalty0
  (9):\penalty0 842--844, 1957.
\newblock \href{https://doi.org/10.1073/pnas.43.9.842}{\ttfamily\path{
  doi:10.1073/pnas.43.9.842}}.

\bibitem[Berge(1958)]{Berge58}
Claude Berge.
\newblock \emph{Th{\'e}orie des graphes et ses applications}.
\newblock Dunod (Orl{\'e}ans, Impr. nouvelle), Paris, 1958.

\bibitem[Harary and Read(1974)]{HararyR74}
Frank Harary and Ronald~C. Read.
\newblock Is the null-graph a pointless concept?
\newblock In \emph{Graphs and Combinatorics}, pages 37--44. Springer, 1974.
\newblock \href{https://doi.org/10.1007/BFb0066433}{\ttfamily\path{
  doi:10.1007/BFb0066433}}.

\end{thebibliography}


\begin{thebibliography}{99}
%
\bibitem[Boers et al (2014)]{Boe14}
Boers, N. and Bookhagen, B. and Barbosa, H. and Marwan, N. and Kurths, J. and Marengo, J. (2014) Prediction of extreme floods in the Eastern Central Andes based on a complex networks approach. \emph{Nature Communications},  \textbf{5(1)},5199, \textbf{URL:~}\emph{10.1038/ncomms6199}.
%
\bibitem[Dirmeyer et al (2012)]{Dir12}
Dirmeyer, P. A. and Cash, B. A. and Kinter, J. L. and et al. (2012) Simulating the diurnal cycle of rainfall in global climate models: Resolution versus parametrization. \emph{Climate Dynamics},  \textbf{39},399–-418, \textbf{DOI:~}\emph{https://doi.org/10.1007/s00382-011-1127-9}
%
\bibitem[Kendon et al (2012)]{Ken12}
Kendon, E.  J. and Roberts, N.  M. and Senior, C. A. and Roberts, M. J . (2012) Realism of Rainfall in a Very High-Resolution Regional Climate Model. \emph{Journal of Climate},  \textbf{25},5791–-5806, \textbf{DOI:~}\emph{https://doi.org/10.1175/JCLI-D-11-00562.1}
%
\bibitem[Goswami et al (2006)]{Gos06}
Goswami, B. N. and Venugopal, V. and Sengupta, D. and Madhusoodanan, M. S. and Xavier, P. K. (2006) Increasing trend of extreme rain events over India in a warming environment. \emph{Science},  \textbf{5804},1442–-1445, \textbf{URL:~}\emph{https://doi.org/10.1126/science.1132027}
%
\bibitem[Mudiar et al. (2018)]{Mud18}
Mudiar, D. and Pawar, S. D. and Hazra, A. and Konwar, M. and Gopalakrishnan, V. and Srivastava, M. K. and Goswami, B. N. (2018) Quantification of observed electrical effect on the raindrop size distribution in tropical clouds. \emph{Journal of Geophysical Research: Atmospheres},  \textbf{123},4527-–4544, \textbf{URL:~}\emph{https://doi.org/10.1029/2017JD028205}.
%
\bibitem[Mudiar et al. (2022)]{Mud22}
Mudiar, D., Hazra, A., Pawar, S. D., Karumuri, R. K., Konwar, M., Mukherjee, S., et al. (2022) Role of electrical effects in intensifying rainfall rates in the tropics. \emph{Geophysical Research Letters},  \textbf{49},e2021GL096276, \textbf{URL:~}\emph{ https://doi.org/10.1029/2021GL096276}.
%
\bibitem[Hazra et al. (2015)]{Haz15}
Hazra, A. and Chaudhari, H. S. and Rao, S. A. and Goswami, B. N. and Dhakate, A. and Pokhrel, S. and Saha, S. K. (2015) Impact of revised cloud microphysical scheme in CFSv2 on the simulation of the Indian summer monsoon. \emph{International Journal of Climatology},  \textbf{35}, 4738-–4755, \textbf{URL:~}\emph{10.1002/joc.4320}.
%
\bibitem[Hazra et al. (2016)]{Haz16}
Hazra, A. and Chaudhari, H. S. and Pokhrel, S. and Saha, S. K. (2016) Indian summer monsoon precipitating clouds: Role of microphysical process rates. \emph{Climate Dynamics},  \textbf{46}, 7--8, \textbf{DOI:~}\emph{https://doi.org/10.1007/s00382-015-2717-8}.
%
\bibitem[Pattnaik et al. (2013)]{Pat13}
Pattnaik, S. and Abhilash, S. and De, S. and Sahai, A. K. and Phani, R. and Goswami, B. N. (2013) Influence of convective parameterization on the systematic errors of Climate Forecast System (CFS) model over the Indian monsoon region from an extended range forecast perspective. \emph{Climate Dynamics},  \textbf{41 (2)}, 341-–365, \textbf{DOI:~}\emph{https://doi.org/10.1007/s00382-013-1662-7}.
%
\bibitem[Saha et al. (2014)]{Saha14}
Saha, S. K. and Pokhrel, S. and Chaudhari, H. S. and Dhakate, A. and Shewale, S. and Sabeerali, C. T. and et al. (2014) Improved simulation of Indian summer monsoon in latest NCEP climate forecast system free run, \emph{International Journal of Climatology},  \textbf{34 (5)}, 1628-1641, \textbf{DOI:~}\emph{10.1002/joc.3791}.
%
\bibitem[Sabeerali et al. (2013)]{Sab13}
Sabeerali, C. T. and Dandi, R. and Dhakate, A. and Salunke, K. and Mahapatra, S. and Rao, S. A. (2013) Simulation of boreal summer intraseasonal oscillations in the latest CMIP5 coupled GCMs, \emph{Journal of Geophysical Research: Atmospheres},  \textbf{118 (21)}, 4401-–4420, \textbf{DOI:~}\emph{10.1002/jgrd.50403}.
%
\bibitem[Goswami et al. (2014)]{Gos14}
Goswami, B. B. and Deshpande, M. and Mukhopadhyay, P. and Saha, S. K. and Rao, S. A. and Murthugudde, R. and Goswami, B. N. (2014) Simulation of monsoon intraseasonal variability in NCEP CFSv2 and its role on systematic bias, \emph{Climate Dynamics},  \textbf{43}, 2725–2745, \textbf{DOI:~}\emph{10.1007/s00382-014-2089-5}.
%
\bibitem[Goswami and Goswami (2017)]{GosGos17}
Goswami, B. B. and Goswami, B. N. (2017) A road map for improving dry-bias in simulating the South Asian monsoon precipitation by climate models, \emph{Climate Dynamics},  \textbf{49}, 2025–2034, \textbf{DOI:~}\emph{10.1007/s00382-016-3439-2}.
%
\bibitem[Hazra et al. (2017)]{Haz17}
Hazra, A. and  Chaudhari, H. S. and Saha, S. K. and Pokhrel, S. and Goswami, B. N. (2017) Progress towards achieving the challenge of Indian Summer Monsoon climate simulation in a Coupled Ocean-Atmosphere Model. \emph{Journal of Advances in Modeling Earth Systems},  \textbf{9(6)},2268-–2290, \textbf{URL:~}\emph{https://doi.org/10.1002/2017MS000966}.
%
\bibitem[Hazra et al. (2020)]{Haz20}
Hazra, A. and Chaudhari, H. S. and Saha, S. K. and Pokhrel, S. and Goswami, B. N. (2020) Role of cloud microphysics in improved simulation of the Asian monsoon quasi-biweekly mode (QBM). \emph{Climate Dynamics},  \textbf{54},599-–614, \textbf{DOI:~}\emph{10.1007/s00382-019-05015-5}.
%
 \bibitem[Pokhrel et al. (2012)]{Pok12}
Pokhrel, S. and Rahaman, H. and Parekh, A. and Saha, S. K. and Dhakate, A. and Chaudhari, H. S. and Gairola, R. M. (2012) Evaporation-precipitation variability over Indian Ocean and its assessment in NCEP Climate Forecast System (CFSv2). \emph{Climate Dynamics},  \textbf{39}, 2585–2608, \textbf{DOI:~}\emph{10.1007/s00382-012-1542-6}.
%
 \bibitem[Saha et al. (2017)]{Saha17}
Saha, S. K. and Sujith, K. and Pokhrel, S. and Chaudhari, H. S. and Hazra, A. (2017) Effects of multilayer snow scheme on the simulation of snow: Offline Noah and coupled with NCEP CFSv2. \emph{Journal of Advances in Modeling Earth Systems},  \textbf{9}, 271-290, \textbf{DOI:~}\emph{10.1002/2016MS00084}.
%
 \bibitem[Pradhan et al. (2022)]{Pra22}
Pradhan, M. and Rao, S. A. and Bhattacharya, A. and Balasubramanian, S. (2022) Improvements in Diurnal Cycle and Its Impact on Seasonal Mean by Incorporating COARE Flux Algorithm in CFS. \emph{Frontiers in Climate},  \textbf{3}, 792980, \textbf{DOI:~}\emph{10.3389/fclim.2021.792980}.
%
\bibitem[Gade et al. (2022)]{Gag22}
Gade, S. V. and Pentakota, S. and Rao, S. A. and Srivastava, A. and Pradhan, M. (2022) Impact of the Ensemble Kalman Filter Based Coupled Data Assimilation System on Seasonal Prediction of Indian Summer Monsoon Rainfall. \emph{Geophysical Research Letters},  \textbf{49}, e2021GL097184, \textbf{DOI:~}\emph{10.1029/2021GL097184}.
%
\bibitem[Abhik et al. (2017)]{Abh17}
Abhik, S. and Krishna, R. P. M. and Mahakur, M. and Ganai, M. and Mukhopadhyay, P. and Dudhia, J. (2017) Revised cloud processes to improve the mean and intraseasonal variability of Indian summer monsoon in climate forecast system: Part 1. \emph{Journal of Advances in Modeling Earth Systems},  \textbf{9}, 1002--1029, \textbf{DOI:~}\emph{10.1002/2016MS000819}.
%
\bibitem[Phani et al. (2023)]{Phan23}
Phani, M. K. R. and Ganai, M. and Tirkey, S. and Mukhopadhyay, P. (2023) Revised cloud processes to improve the simulation and prediction skill of Indian summer monsoon rainfall in climate forecast system model. \emph{Climate Dynamics}, \textbf{DOI:~}\emph{10.1007/s00382-023-06674-1}.
%
\bibitem[Ganai et al. (2019)]{Gana19}
Ganai, M. and Mukhopadhyay, P. and Phani, M. K. R. and Abhik, S. and Halder, M. (2019) Revised cloud and convective parameterization in CFSv2 improve the underlying processes for northward propagation of Intraseasonal oscillations as proposed by the observation-based study PDF file, \emph{Climate Dynamics},  \textbf{53}, 2793–2805, \textbf{DOI:~}\emph{10.1007/s00382-019-04657-9}.
%
\bibitem[Pruppacher and Klett (2010)]{PruKle10}
Pruppacher,  H. R. and Klett, J. D. (2010) Microphysics of Clouds and Precipitation. \emph{Springer Science and Business Media}, \textbf{DOI:~}\emph{https://doi.org/10.1007/978-0-306-48100-0}.
%
\bibitem[Wu et al. (2018)]{WuXiDonZha18}
Wu, P. and Xi, B. and Dong, X. and Zhang, Z. (2018) Evaluation of autoconversion and accretion enhancement factors in general circulation model warm-rain parameterizations using ground-based measurements over the Azores. \emph{Atmospheric Chemistry and Physics},  \textbf{18 (23)}, 17405–-17420, \textbf{DOI:~}\emph{https://doi.org/10.5194/acp-18-17405-2018}.
%
\bibitem[Liu et al. (2004)]{LiuDauMc04}
Liu, Y. and Daum, P. H. and McGraw, R. (2004) An analytical expression for predicting the critical radius in the autoconversion parameterization. \emph{Geophysical Research Letters},  \textbf{31}, L06121, \textbf{DOI:~}\emph{10.1029/2003GL019117}.
%
\bibitem[Liu et al. (2006)]{LiuDauMc06}
Liu, Y. and Daum, P. H. and McGraw, R. (2006) Parameterization of the Autoconversion Process. Part II: Generalization of
Sundqvist-Type Parameterizations. \emph{Journal of the Atmospheric Sciences},  \textbf{63}, 1103-–1109, \textbf{DOI:~}\emph{10.1175/JAS3675.1}.
%
\bibitem[Kessler (1969)]{Kes69}
Kessler, E. (1969) On the distribution and continuity of water substance in atmospheric circulation. \emph{Meteorological Monographs}, \textbf{32}, 84, \textbf{DOI:~}\emph{10.1007/978-1-935704-36-2\_1}.
%
\bibitem[Sundqvist (1978)]{Sun78}
Sundqvist, H. (1978) A parameterization scheme for non-convective condensation including prediction of cloud water content. \emph{Quarterly Journal of the Royal Meteorological Society},  \textbf{104}, 677-690, \textbf{DOI:~}\emph{ https://doi.org/10.1002/qj.49710444110}.
%
\bibitem[Zhao and Carr (1997)]{ZaoCar97}
Zhao, Q. and Carr, F. H. (1997) A prognostic cloud scheme for operational NWP models. \emph{Monthly Weather Review},  \textbf{125}, 1931-–1953, \textbf{DOI:~}\emph{10.1175/1520-0493(1997)125<1931:apcsfo>2.0.co;2}. 
%
\bibitem[Han and Pan (2011)]{HanPan11}
Han, J. and Pan, H. L. (2011) A prognostic cloud scheme for operational NWP models. \emph{Weather and Forecasting},  \textbf{26 (4)}, 520-–533, \textbf{DOI:~}\emph{10.1175/WAF-D-10-05038.1}. 
%
\bibitem[Stocker et al. (2001)]{Sto01}
Stocker, T. and Clarker, G. K. C. and Treut, H. L. and et al. (2001) Physical climate processes and feedbacks. \emph{Climate Change 2001: The Scientific Basis, J. T. Houghton et al., Eds., Cambridge University Press},  419-–470. 
%
\bibitem[Bhowmik et al. (2023a)]{Bhow23a}
Bhowmik, M. and Hazra, A. and Rao, S. A. and Wang, L. P. (2023a) Eulerian-Lagrangian particle-based model for diffusional growth for the better parameterization of ISM clouds: A road map for improving climate model through small-scale model using observations. \emph{Preprint}.
%
\bibitem[Cheng et al. (2007)]{CheWanChe07}
Cheng, C. T. and  Wang, W. C. and Chen, J. P. (2007) A modeling study of aerosol impacts on cloud microphysics and radiative properties. \emph{Quarterly Journal of the Royal Meteorological Society},  \textbf{133}, 283–-297, \textbf{DOI:~}\emph{10.1002/qj.25}.
%
\bibitem[Cheng et al. (2010)]{CheWanChe10}
Cheng, C. T. and  Wang, W. C. and Chen, J. P. (2010) Simulation of the effects of increasing cloud condensation nuclei on mixed-phase clouds and precipitation of a front system. \emph{Atmospheric Research},  \textbf{96}, 461--476, \textbf{DOI:~}\emph{10.1016/j.atmosres.2010.02.005}.
%
\bibitem[Thompson et al. (2008)]{ThomFieRas08}
Thompson, G. and Field, P. R. and Rasmussen, R. M. and Hall, W. D. (2009) Explicit forecasts of winter precipitation using and improved bulk microphysics scheme. Part II: Implementation of a new snow parameterization. \emph{Monthly Weather Review},  \textbf{136}, 5095–-5115, \textbf{DOI:~}\emph{10.1175/2008MWR2387.1}.
%
\bibitem[Wu et al. (2010)]{WuYu10}
Wu, T. and Yu, R. and Zhang, F. and Wang, Z. and Dong, M. and Wang, L. and Jin, X. and Chen, D. and Li, L. (2010) The Beijing Climate Center atmospheric general circulation model: Description and its performance for the present-day climate. \emph{Monthly Weather Review},  \textbf{34}, 123-–147, \textbf{DOI:~}\emph{10.1007/s00382-008-0487-2}.
%
\bibitem[Martin et al. (1994)]{MarJohSpi94}
Martin, G. M. and Johnson, D. W. and Spice, A. (1994) The measurement and parameterization of effective radius of droplets in warm stratocumulus clouds. \emph{Journal of the Atmospheric Sciences},  \textbf{51}, 1823--1842, \textbf{DOI:~}\emph{10.1175/1520-0469(1994)051<1823:TMAPOE>2.0.CO;2}.
%
\bibitem[Lohmann et al. (2007)]{Loh07}
Lohmann, U. and Stier, P. and  Hoose, C. and Ferrachat, S. and Kloster, S. and Roeckner, E. and Zhang, J. (2007) Cloud microphysics and aerosol indirect effects in the global climate model ECHAM5-HAM. \emph{Atmospheric Chemistry and Physics},  \textbf{7 (13)}, 3425-–3446, \textbf{DOI:~}\emph{ https://doi.org/10.5194/acp-7-3425-2007}.
%
\bibitem[Morrison and Gettelman (2008)]{MorGet08}
Morrison, H. and Gettelman, A. (2008) A new two-moment bulk stratiform cloud microphysics scheme in the community atmosphere model, version 3 (CAM3). Part I: Description and numerical tests. \emph{Journal of Climate},  \textbf{21}, 3642–-3659, \textbf{DOI:~}\emph{https://doi.org/10.1175/2008JCLI2105.1}.
%
\bibitem[Sundqvist et al. (1989)]{Sun89}
Sundqvist, H. and Berge, E. and Kristjánsson, J. E. (1989) Condensation and Cloud Parameterization Studies with a Mesoscale Numerical Weather Prediction Model. \emph{Monthly Weather Review}, \textbf{117(8)}, 1641–-1657, \textbf{DOI:~}\emph{https://doi.org/10.1175/1520-0493(1989)117<1641:CACPSW>2.0.CO;2}.
%
\bibitem[Dutta et al. (2021)]{Dutt21}
Dutta, U. and Hazra, A. and Chaudhari, H. S. and Saha, S. K.and Pokhrel, S. and  Shiu, C. J. and Chen J. P. (2021) Role of microphysics and convective autoconversion for the better simulation of tropical intraseasonal oscillations (MISO and MJO). \emph{Journal of Advances in Modeling Earth Systems},  \textbf{13}, e2021MS002540, \textbf{DOI:~}\emph{ 10.1029/2021MS002540}.
%
\bibitem[Rotstayn and Liu (2005)]{RotLiu05}
Rotstayn, L. D. and Liu, Y. (2005) A smaller global estimate of the second indirect aerosol effect. \emph{Geophysical Research Letters},  \textbf{32}, L05708, \textbf{DOI:~}\emph{10.1029/2004GL021922}.
%
\bibitem[Rasch and Kristjánsson (1998)]{RasKri98}
Rasch, P. J. and Kristjánsson, J. E. (1998) A Comparison of the CCM3 Model Climate Using Diagnosed and Predicted Condensate Parameterizations. \emph{Journal of Climate},  \textbf{11(7)}, 1587–-1614, \textbf{DOI:~}\emph{10.1175/1520-0442(1998)011<1587:ACOTCM>2.0.CO;2}.
%
\bibitem[Rotstayn (2000)]{Rot00}
Rotstayn, L. D. (2000) On the “tuning” of autoconversion parameterizations in climate models. \emph{Journal of Geophysical Research},  \textbf{105}, 15495--15507, \textbf{DOI:~}\emph{10.1029/2000JD900129}.
%
\bibitem[Lamb and Verlinde (2011)]{LamVer11}
Lamb, D. and Verlinde, J. (2011) Physics and Chemistry of Clouds. \emph{Cambridge University Press,Cambridge}, \textbf{DOI:~}\emph{10.1017/CBO9780511976377}.
%
\bibitem[Rao et al. (2019)]{Rao19}
Rao, S. A. and Goswami, B. N. and Sahai, A. K. and Rajagopal, E. N. and Mukhopadhyay, P. and Rajeevan, M. and Nayak, S. and et al. (2023b) Monsoon Mission : A targeted activity to improve monsoon prediction across scales. \emph{Bulletin of the American Meteorological Society}, \textbf{100}, 2509--2532, \textbf{DOI:~}\emph{10.1175/BAMS-D-17-0330.1}.
%
\bibitem[Kumar et al. (2014)]{KumHaz14}
Kumar, S. and Hazra, A. and Goswami, B. N. (2014) Role of interaction betweendynamics, thermodynamics and cloud microphysics on summer mon-soon precipitating clouds over the Myanmar coast and the WesternGhats, \emph{climate Dynamics}, \textbf{43}, 911--924, \textbf{DOI:~}\emph{10.1007/s00382-013-909-3}.
%
\bibitem[Krishnamurti et al. (1989)]{Kri89}
Krishnamurti, T. N. and Bedi, H. S. and Subramaniam, M. (1989) The Summer Monsoon of 1987, \emph{Journal of Climate}, \textbf{2}, 321–-340, \textbf{DOI:~}\emph{10.1175/1520-0442(1989)002<0321:TSMO>2.0.CO;2}.
%
\bibitem[Murakami (1980)]{Mur80}
Murakami, T. (1980) Temporal variations of satellite-observed outgoing longwave radiation over the winter monsoon region. Part II: short-period (4-6 day) oscillations, \emph{Monthly Weather Review}, \textbf{108}, 427-–444, \textbf{DOI:~}\emph{10.1175/1520-0493(1980)108<0427:TVOSOO>2.0.CO;2}.
%
\bibitem[Chaudhari et al. (2010)]{Chau10}
Chaudhari, H. S. and Shinde, M. A. and Oh, J. H. (2010) Understanding of anomalous Indian Summer Monsoon rainfall of 2002 and 1994, \emph{Quaternary International}, \textbf{213}, 20--32, \textbf{DOI:~}\emph{10.1016/j.quaint.2008.05.009}.
%
\bibitem[Prasad and Verma (1985)]{Praver85}
Prasad, K. D. and Verma, R. K. (1985) Large-scale features of satellite-derived outgoing long-wave radiation in relation to monsoon circulation over the indian region, \emph{Journal of Climatology}, \textbf{5(3)}, 297–-306, \textbf{DOI:~}\emph{10.1002/joc.3370050306}.
%
\bibitem[Chaudhari et al. (2019)]{Cha19}
Chaudhari, H.S. and Hazra, A. and Pokhrel, S. and Saha, S. K. and Taluri, S.S. (2019) Simulation of extreme Indian summer monsoon years in Coupled Model Intercomparison Project Phase 5 models Role of cloud processes, \emph{International Journal of Climatology}, \textbf{39}, 901--920, \textbf{DOI:~}\emph{10.1002/joc.5851}.
%
\bibitem[Liebmann and Smith (1996)]{LieSmi96}
Liebmann, B. and Smith, C. A. (1996) Description of a Complete (Interpolated) Outgoing Longwave Radiation Dataset, \emph{Bulletin of the American Meteorological Society}, \textbf{77}, 1275--1277, \textbf{DOI:~}\emph{http://www.jstor.org/stable/26233278}.
%
\bibitem[Morrison et al. (2020)]{Mor20}
Morrison, H. and Lier-Walqui, M. and Fridlind, A. M. and Grabowski, W. W. and Harrington, J. Y. and Hoose, C. and et al. (2020) Confronting the challenge of modeling cloud and precipitation microphysics. \emph{Journal of Advances in Modeling Earth Systems},  \textbf{12},e2019MS001689, \textbf{DOI:~}\emph{https://doi.org/10.1029/2019MS001689}.
%
\bibitem[Grabowski and Wang (2013)]{GraWang13}
Grabowski, W. W. and Wang, L. P. (2013) Growth of cloud droplets in a turbulent environment. \emph{Annual Review of Fluid Mechanics},  \textbf{45}, 293--324, \textbf{DOI:~}\emph{https://doi.org/10.1146/annurev-fluid-011212-140750}.
%
\bibitem[Grabowski et al. (2019)]{Grab19}
Grabowski, W. W. and Morrison, H. and Shima, S. I. and Abade, G. and Pawlowska, H. and Dziekan, P. (2019) Modeling of cloud microphysics: Can we do better?. \emph{Bulletin of the American Meteorological Society},  \textbf{100},655-–672, \textbf{URL:~}\emph{https://doi.org/10.1175/bams-d-18-0005.1}.
%
\bibitem[Chen et al. (2020)]{Chen20}
Chen, S. and Xue, L. and Yau, M. K. (2020) Impact of aerosols and turbulence on cloud droplet growth: An in-cloud seeding case study using a parcel–DNS (direct numerical simulation) approach. \emph{Atmospheric Chemistry and Physics },  \textbf{20}, 10111--10124 \textbf{DOI:~}\emph{https://doi.org/10.5194/acp-20-10111-2020}.
%
\bibitem[Hazra et al. (2013)]{Haz13}
Hazra, A. and Goswami, B. N. and Chen, J. P. (2013) Role of interactions between aerosol radiative effect, dynamics, and cloud microphysics on transitions of monsoon intraseasonal oscillations. \emph{Journal of the Atmospheric Sciences},  \textbf{70}, 2073--2087 \textbf{DOI:~}\emph{10.1175/JAS-D-12-0179.1}.
%
\end{thebibliography}
\end{document}